\documentclass[a4paper,11pt]{article}
\pdfoutput=1
\usepackage{jcappub}
\usepackage{threeparttable} 
\usepackage[T1]{fontenc}
\usepackage{color}
\usepackage{subfigure}

\DeclareFontFamily{OT1}{rsfs}{} 
\DeclareFontShape{OT1}{rsfs}{m}{n}{<-7> rsfs5 
    <7-10> rsfs7 <10-> rsfs10}{}   
\DeclareMathAlphabet{\scr}{OT1}{rsfs}{m}{n}

\newcommand{\ham}{\scr{H}}
\newcommand{\Mp}{m_\text{Pl}}
\newcommand{\M}{M_{\star}}
\newcommand{\di}{\partial}
\newcommand{\epsilonH}{\epsilon_{\text{H}}}
\newcommand{\etaH}{\eta_{\text{H}}}
\newcommand{\meff}{m_\text{eff}}
\renewcommand{\k}{\mathbf{k}}
\newcommand{\x}{\mathbf{x}}
\newcommand{\hamk}{\,\,\,\hat{\!\!\!\scr{H}}_{\!\!\k}}

\title{Primordial fluctuations from deformed quantum algebras}

\author[a,b]{Andrew C.\ Day,}
\author[c]{Iain A.\ Brown}
\author[b,1]{and Sanjeev S.\ Seahra\note{Corresponding author.}}

\affiliation[a]{Department of Applied Mathematics, University of Western Ontario \\ London, ON, Canada N6A 5B7}
\affiliation[b]{Department of Mathematics and Statistics, University of New Brunswick \\ Fredericton, NB, Canada E3B 5A3}
\affiliation[c]{Institute of Theoretical Astrophysics, University of Oslo \\ P.O. Box 1029 Blindern, N-0315 Oslo, Norway}

\emailAdd{aday46@uwo.ca}
\emailAdd{ibrown@astro.uio.no}
\emailAdd{sseahra@unb.ca}

\abstract{We study the implications of deformed quantum algebras for the generation of primordial perturbations from slow-roll inflation.  Specifically, we assume that the quantum commutator of the inflaton's amplitude and momentum in Fourier space gets modified at energies above some threshold $M_{\star}$.  We show that when the commutator is modified to be a function of the momentum only, the problem of solving for the post-inflationary spectrum of fluctuations is formally equivalent to solving a one-dimensional Schr\"odinger equation with a time dependent potential.  Depending on the class of modification, we find results either close to or significantly different from nearly scale invariant spectra.  For the former case, the power spectrum is characterized by step-like behaviour at some pivot scale, where the magnitude of the jump is $\mathcal{O}(H^{2}/M_{\star}^{2})$. ($H$ is the inflationary Hubble parameter.)  We use our calculated power spectra to generate predictions for the cosmic microwave background and baryon acoustic oscillations, hence demonstrating that certain types of deformations are incompatible with current observations.}

\begin{document}
\maketitle
\flushbottom

\section{Introduction}

One of the most pressing issues confronting any quantum theory of gravitation is how it can be verified or falsified using real world experiments and observations.  Because such theories naturally involve Planckian energy scales $10^{19}\,\text{GeV}$ or lengths $10^{-35}\,\text{m}$, there are very few known phenomenon that can realistically exhibit observable quantum gravity effects.  The most promising possibility is the cosmic microwave background (CMB) and large scale structure of the universe, both of which are sensitive to the spectrum of density fluctuations in the presumptive inflationary era (when temperature of the universe was a few orders of magnitude less than the Planck scale).  Indeed, inflation is a natural ``microscope'' with which we can test exotic small scale physics:  Exponential expansion means that if one traces the size of present day sub-horizon scale fluctuations backwards in time, one finds that they have sub-Planckian wavelengths at some point during inflation.  Hence, any process governing the generation and evolution of cosmological during inflation should be sensitive to very small scale physics.

In the absence of a full and calculable quantum theory of gravity, it is useful to consider various classes of phenomenological effects and how they might manifest them in the primordial perturbative spectrum.  In the literature, various authors have considered modifications to scalar field dispersion relations \cite{Brandenberger:2000wr,Martin:2000xs,Niemeyer:2000eh,Shankaranarayanan:2002ax}, non-commutativity \cite{Chu:2000ww,Lizzi:2002ib,Brandenberger:2002nq,Hassan:2002qk}, or modified uncertainty relations \cite{Kempf:2000ac,Easther:2001fi,Kempf:2001fa,Ashoorioon:2004vm,Kempf:2006wp}.  There have also been attempts also to calculate trans-Planckian contributions to the primordial power spectrum in a model-independent way by imposing initial conditions on a ``new physics hyeprsurface'' \cite{Niemeyer:2002kh,Bozza:2003pr,Easther:2002xe,Danielsson:2002kx}.  Recently, effects from Horava-Lifshitz gravity \cite{Ferreira:2012xa} and polymer quantization \cite{Seahra:2012un} have been reported.  A feature of many (but not all) of these studies is that short distance effects induce changes to the conventional scale-invariant power spectrum with amplitude $(H/\M)^{\gamma}$, where $\M$ is the energy threshold above which the modifications are important, $H$ is the inflationary Hubble parameter, and the power $\gamma \gtrsim 1$ depends on the model.

In this paper, we consider the possibility that quantum algebra of phase space variables gets deformed at high energy.  That is, we modify the standard prescription that canonical Poisson brackets $\{q,p\} = 1$ get mapped to commutators $[\hat{q},\hat{p}] = i$ when a system is quantized.  Rather, we take the quantum algebra to be
\begin{equation}
[\hat{q},\hat{p}] = if(\beta\hat{p}),
\end{equation}
where $\beta$ is dimensionful parameter that defines an energy threshold and $f$ is a function satisfying some mild assumptions.\footnote{Interestingly, one of the first people to suggest this type of modification was Heisenberg himself \cite[see references therein]{2013arXiv1301.0116B}.} The implications of deformed quantum algebras (DQAs) have been investigated for the simple harmonic oscillator \cite{KMM,2013arXiv1301.0116B} and quantum field theory \cite{Berger:2010pj,Husain:2012im}.  Here, we follow references \cite{Berger:2010pj,Husain:2012im} and assume that the commutator between the Fourier components of scalar field amplitude and momentum receives high energy corrections, and hence derive a modified spectrum of primordial perturbations. 

The paper is organized as follows:  In section \ref{sec:classical}, we review the gauge-invariant classical theory of density perturbations during slow-roll inflation.  In section \ref{sec:quantization}, we show how to quantize the perturbations with generic DQAs, and with specific choices $f(x) = 1$ and $f(x) = 1 \pm x^{2}$.   In the same section, we also give our generalization of the Bunch-Davies initial conditions situations where $f(x) \ne 1$, and explicitly show that once can find the primordial power spectrum at the end of inflation by solving a Schr\"odinger equation with a time varying potential.  In section \ref{sec:qualitative}, we discuss a few analytical properties and qualitative features of the power spectra for the specific cases of section \ref{sec:quantization}.  In section \ref{sec:results}, we present our numerically calculated power spectra and the associated predictions for the CMB and baryon acoustics oscillations.  For the case $f(x) = 1+x^{2}$, we see $\mathcal{O}(H^{2}/\M^{2})$ DQA effects on power spectra on small scales. On the other hand, we see significant tension between the $f(x) = 1 - x^{2}$ case and standard results, likely indicating that this choice can only be made to match observations for extreme parameter choices.  Section \ref{sec:discussion} is reserved for conclusions, while appendices \ref{sec:eigenvalues} and \ref{sec:numeric} give technical details about out analytic and numeric calculations, respectively.

\section{Formalism}\label{sec:formalism}

\subsection{Classical equations}\label{sec:classical}

We will consider single field inflation with action
\begin{equation}\label{eq:action 1}
	S = \int d^{4}x \sqrt{-g} \left[ \frac{1}{2}\Mp^{2} R - \frac{1}{2} g^{\alpha\beta} \di_{\alpha} \phi \, \di_{\beta} \phi - V(\phi) \right],
\end{equation}
with the reduced Planck mass $\Mp^{2} = 1/8\pi G$.  As usual, we consider perturbations about an FRW background
\begin{equation}
	ds^{2} = -dt^{2} + a^{2}(t) d\mathbf{x}^{2} = a^{2}(\eta) (-d\eta^{2} + d\mathbf{x}^{2}),
\end{equation}
where the evolution of the homogeneous scalar field $\phi = \phi(t)$ and the Hubble parameter $H = H(t) = \dot{a}/a$ is given by
\begin{equation}
	0 = \ddot{\phi} + 3H\dot\phi + \di_{\phi}V, \quad \dot{H} = -\dot{\phi}^{2}/2\Mp^{2}.
\end{equation}
We perturb the scalar field and the spatial curvature of metric as follows:
\begin{align}
	\delta\phi = \delta\phi(t,\mathbf{x}), \quad \delta R^{(3)} = -(4/a^{2}) \Delta^{(3)} \mathcal{R}(t,\mathbf{x}) ,
\end{align}
where $\Delta^{(3)} = \delta^{ij} \di_{i} \di_{j} = \mathbf{\nabla}^{2}$ with $i,j=1,2,3$.  Putting these perturbations into the original action (\ref{eq:action 1}), we find the second order contribution to $S$ from $\varphi$ and $\mathcal{R}$ to be
\begin{equation}\label{eq:action 2}
	S_{2}[\chi] = \int d^{4}x \sqrt{-g} \left[ -  \frac{1}{2} g^{\alpha\beta} \di_{\alpha} \chi \di_{\beta} \chi -   \frac{1}{2} \meff^{2} \chi^{2} \right],
\end{equation}
where
\begin{equation}
	\meff^{2}  = - \frac{H}{a^{3}\dot\phi} \frac{d}{dt} \left[ a^{3} \frac{d}{dt} \left(\frac{\dot\phi}{H} \right) \right], \quad
	\chi = \delta\phi - \frac{\dot\phi}{H} \mathcal{R}. 
\end{equation} 
The field $\chi$ is gauge invariant and equal to the perturbation to the scalar field amplitude on flat $\mathcal{R} = 0$ slicings.  Also, it is related to the ubiquitous Mukhanov-Sasaki \cite{Mukhanov:1985rz,Sasaki:1986hm} variable by $v = a\chi$.   In terms of $\chi$, the comoving curvature perturbation is
\begin{equation}
	\mathcal{R}_{\text{c}} = - \frac{H}{\dot{\phi}} \chi.
\end{equation}
The time dependent effective mass in (\ref{eq:action 2}) can be expressed as
\begin{equation}
	\frac{\meff^{2}}{H^{2}} = 3\etaH - 3\epsilonH - \etaH^{2} - \epsilonH \etaH + 2\epsilonH^{2} + \frac{1}{H} \frac{d\etaH}{dt} - \frac{2}{H} \frac{d\epsilonH}{dt},
\end{equation}
where the (Hamiltonian-Jacobi) slow-roll parameters are given by \cite{Lyth:2009zz}
\begin{equation}
	\epsilonH = -\frac{\dot{H}}{H^{2}},  \quad  \etaH = -\frac{\ddot{H}}{2H\dot{H}}.
\end{equation}
Hence, in slow-roll inflation we have $\meff^{2} \ll H^{2}$.

We now introduce the Fourier transform of $\chi$ as follows:
\begin{equation}
\chi(t,\x) =  \frac{1}{\sqrt{V_{0}}}\sum_{\k}{\tilde\chi}_{\k}(t) e^{i {\k}\cdot{\x}}, \quad
{\tilde\chi}_\k(t) = \frac{1}{\sqrt{V_{0}}}\int d^3x\ e^{-i\k\cdot \x} \chi(t,\x), \quad V_{0} = \int d^{3}x.
\end{equation}
Here, $V_{0}$ is the volume associated with our box normalization.  After a suitable re-labelling of the Fourier coefficients and removal of redundant degrees of freedom \cite{Mahajan:2007qg}, the action becomes
\begin{equation}\label{eq:action 3}
	S_{2}[\chi] = \sum_{\k} \int dt \, a^{3}\left[ \frac{1}{2} \dot\chi_{\k}^{2} -   \frac{1}{2} \left( \frac{k^{2}}{a^{2}} + \meff^{2} \right) \chi_{\k}^{2} \right],
\end{equation}
where $\chi_{\k} \in \mathbb{R}$.  This is essentially the action of a collection of decoupled harmonic oscillators with time dependent masses and frequencies.  The Hamiltonian is
\begin{equation}
	\ham = \sum_{\k} \ham_{\k}, \quad \ham_{\k} = \frac{1}{2a^{3}} \pi_{\k}^{2} + \frac{a^{3}}{2} \left( \frac{k^{2}}{a^{2}} + \meff^{2} \right) \chi_{\k}^{2},
\end{equation}
where $\pi_{\k}$ is the momentum conjugate to $\chi_{\k}$ such that
\begin{equation}
	\{ \chi_{\k}, \pi_{\k'} \} = \delta_{\k,\k'}.
\end{equation}

\subsection{Quantization with deformed quantum algebras}\label{sec:quantization}

We now consider the quantization of the system described by the action (\ref{eq:action 3}).  Our ultimate goal is the power spectrum of the comoving curvature perturbation, which is defined by
\begin{equation}\label{eq:spectrum}
	\mathcal{P}_{\mathcal{R}_{c}}(k) = \frac{k^{3}}{2\pi^{2}} \left[ \frac{H^{2}}{\dot\phi^{2}} \langle \chi_{\k}^{2} \rangle \right]_{k \ll Ha}.
\end{equation}
Here, the expectation value $\langle \chi_{\k}^{2} \rangle$ is to be evaluated in some suitably defined ``vacuum state''.  We will work to leading order in slow roll parameters, which means that we can consistently calculate $\langle \chi_{\k}^{2} \rangle$ in the de Sitter approximation; i.e., we set $\meff = 0$ and $a=\exp(Ht)$ in (\ref{eq:action 3}). 

The structure of the Hamiltonian implies the quantum state of the system will be given as a tensor product:
\begin{equation}
	| \psi \rangle = \bigotimes_{\k} |\psi_{\k} \rangle.
\end{equation}
In the Schr\"odinger picture, the quantum state vector associated with a given $\k$ satisfies the evolution equation:
\begin{equation}\label{eq:schrodinger 1}
	i \frac{d}{dt} |\psi_{\k} \rangle = \hamk | \psi_{\k}\rangle .
\end{equation} 
We consider the following class of deformed quantum algebras (DQAs):
\begin{equation}\label{eq:commutator}
	[ \hat\chi_{\k}, \hat\pi_{\k'}]=i\delta_{\k,\k'} f \left( \frac{\hat\pi_{\k}}{\M^{1/2}a^{3/2}} \right).
\end{equation} 
Here, $\M$ is the ``DQA energy scale'' and the smooth even function $f$ satisfies $f(0) = 1$, but is otherwise arbitrary.  The $f(0)=1$ condition allows us to recover the standard commutator in the limit $\M \rightarrow \infty$, while the fact that $f$ is even ensures that the commutator does not depend on the sign of $\hat\phi_{\k}$.  We will see that $\M$ represents the energy threshold of exotic physics in our formalism.  The operator algebra (\ref{eq:commutator}) is similar to the one considered in \cite{Husain:2012im}, but with an additional factor of $a^{3/2}$ in the argument of $f$.  This arises from the behaviour of various quantities under a change of scale $\mathbf{x} \rightarrow \gamma \mathbf{x}$:
\begin{equation}
	a \rightarrow \gamma^{-1} a, \quad \hat{\chi}_{\k} \rightarrow \gamma^{3/2} \hat{\chi}_{\k}, \quad \hat{\pi}_{\k} \rightarrow \gamma^{-3/2} \hat{\pi}_{\k}.
\end{equation}
We see that the $a^{3/2}$ factor is exactly what is needed to make the modified commutator (\ref{eq:commutator}) invariant under such a transformation.

Let us introduce a basis $|z\rangle$ labelled by the continuous real parameter $z \in [z_{-},z_{+}]$.  We write quantum states as wavefunctions:
\begin{equation}
	\psi(t,z) = \langle z | \psi_{\k} \rangle, \quad \psi \in L^{2}([z_{-},z_{+}],dz), \quad \psi(t,z_{\pm}) = 0.
\end{equation} 
In order to reproduce (\ref{eq:commutator}), we define the actions of $\hat\chi_{\k}$ and $\hat\pi_{\k}$ in this basis as:
\begin{subequations}
\begin{align}
	\langle z | \hat\chi_{\k} | \psi_{\k} \rangle &  = i \di_{z} \psi(t,z), \\ \label{eq:momentum operator} \langle z | \hat{\pi}_\k | \psi_{\k} \rangle & = \M^{1/2}a^{3/2} U(z/\M^{1/2}a^{3/2}) \psi(t,z).
\end{align}
\end{subequations}
Here, the function $U$ satisfies the differential equation:
\begin{equation}\label{eq:U def}
	U'(x) = f(U(x)), \quad U(0)=0.
\end{equation}
In this representation, the Schr\"odinger equation (\ref{eq:schrodinger 1}) becomes
\begin{equation}\label{eq:schrodinger 2}
	i \frac{\di \psi}{\di t} = \frac{1}{2} \left[ \M  U^{2} \left( \frac{z}{\M^{1/2}a^{3/2}} \right) - k^{2}a \frac{\di^{2}}{\di z^{2}} \right] \psi.
\end{equation}
We demand that (\ref{eq:momentum operator}) define a unique mapping from $z$ to $\pi_{\k}$, hence we must choose $z_{\pm}$ such that $U$ is single valued on $[z_{-},z_{+}]$. The appropriate inner product between wavefunctions is
\begin{equation}
	\langle \tilde\psi_{\k} | \psi_{\k} \rangle = \int_{z_{-}}^{z_{+}} dz \, \tilde\psi^{*} \psi,
\end{equation}
and the quantity that we need to obtain the spectrum (\ref{eq:spectrum}) is 
\begin{equation}
	\langle \chi_{\k}^{2} \rangle = -\int_{z_{-}}^{z_{+}} dz\,\psi^{*} \frac{\di^{2}}{\di z^{2}} \psi =  \int_{z_{-}}^{z_{+}} dz\, \left| \frac{\di\psi}{\di z} \right|^{2}. 
\end{equation} 

Equation (\ref{eq:schrodinger 2}) has complicated time dependence which can be somewhat simplified by changing coordinates:
\begin{equation}
	\eta = -\frac{1}{Ha}, \quad y = \frac{z}{\sqrt{k} a},
\end{equation} 
and rescaling the wavefunction
\begin{equation}\label{eq:wavefunction re-scaling}
	\psi(t,z) = \frac{1}{k^{1/4}a^{1/2}} \varphi(\eta,y) \exp\left( - \frac{iy^{2}}{2k\eta} \right).
\end{equation}
In terms of these quantities, the Schr\"odinger equation becomes
\begin{equation}\label{eq:schrodinger 3}
	i \frac{\di \varphi}{\di \eta} = \mathcal{A} \varphi, \quad \mathcal{A} =  \frac{k}{2} \left[ \frac{U^{2}(\sqrt{g}y)}{g} - \frac{\di^{2}}{\di y^{2}} \right], \quad \varphi(\eta,y_\pm) = 0,
\end{equation}
where $y_\pm = z_{\pm}/\sqrt{k}a$. Solutions of this equation are elements of $L^{2}([y_-,y_+],dy)$ and we have the inner product
\begin{equation}\label{eq:inner product}
	\langle \tilde\varphi | \varphi \rangle = \int_{y_-}^{y_+} dy \, \tilde\varphi^{*} \varphi.
\end{equation}
In (\ref{eq:schrodinger 3}), the time-dependent quantity $g$ is given by
\begin{equation}
	g = \frac{k}{\M a} = -k\eta\epsilon, \quad \epsilon = \frac{H}{\M}.
\end{equation}
We call $\epsilon$ the ``adiabaticity parameter'' since
\begin{equation}
	\frac{d\mathcal{A}}{d\eta} \propto \epsilon;
\end{equation}
i.e., $\epsilon$ controls how fast the time-dependent potential in (\ref{eq:schrodinger 3}) varies.  Also, we call $g$ the ``DQA coupling'' parameter since it can be re-written as
\begin{equation}
	g = \frac{\lambda_\star}{\lambda_\text{phys}}, \quad \lambda_\text{phys} = \frac{2\pi a}{k}, \quad \lambda_\star = \frac{2\pi}{M_\star},
\end{equation}
where $\lambda_\text{phys}$ is the physical wavelength of the mode and $\lambda_\star$ is the Compton wavelength of the DQA energy scale $M_{\star}$.   Intuitively, we expect the effects of DQAs to be small in the small coupling $g \ll 1$ regime since large wavelength modes ought to be insensitive to exotic short distance physics.  Conversely, we expect to see significant changes to the quantum state evolution in the large coupling $g \gg 1$ (or small wavelength) regime.     

In order to fix the time evolution of the quantum state, we need to specify initial data for the wavefunction.  We will see in \S\ref{sec:case 1} that the standard Bunch-Davies vacuum defined for the conventional $f(x) = 1$ commutator is the quantum state that minimizes $\langle \mathcal{A} \rangle$ at the start of inflation.\footnote{Note, this is not the same thing as demanding that our initial data minimizes the expectation value of the Hamiltonian $\langle \hamk \rangle$.  The reason for the discrepancy is due to the $y$-dependent phase factor in the wavefunction re-scaling (\ref{eq:wavefunction re-scaling}).  However, in the subhorizon limit $k/Ha \gg 1$ this phase is negligible and our initial data will be an excellent approximation to the ground state of the Hamiltonian.}    We adopt an identical identical initial data prescription when $f(x) \ne 1$.  Stated another way, suppose that we can solve the eigenvalue problem
\begin{equation}\label{eq:eigenvalue problem}
	E_{n} \varphi_{n} = \mathcal{A} \varphi_{n}, \quad \langle \varphi_{n} | \varphi_{m} \rangle = \delta_{n,m}.
\end{equation}
Bunch-Davies like initial conditions correspond to setting 
\begin{equation}\label{eq:Bunch-Davies ICs}
	\varphi\big{|}_{\eta = \eta_{0}} = \varphi_{0},
\end{equation}
where $\varphi_{0}$ is the eigenfunction of $\mathcal{A}$ associated with the smallest eigenvalue $E_{0}$, and $\eta=\eta_{0}$ is the start of inflation.  Note that the time dependence of $\mathcal{A}$ implies that if we prepare the system in its ``instantaneous'' ground state initially, it will not generally stay in that ground state as time progresses.  The exceptions to this conclusion are when $f(x)=1$ or when one enforces the adiabatic limit $\epsilon \rightarrow 0$; in either case, the time dependence of $\mathcal{A}$ drops out and the mode will stay in its ground state.

In terms of $y$ and $\varphi$, the expectation value of the field amplitude is:
\begin{equation}\label{eq:chi squared}
	\langle \chi_{\k}^{2} \rangle = \frac{H^{2}}{k^{3}} \int_{y_-}^{y_+} dy\, \left\{ y^{2} |\varphi|^{2} + ( -k\eta)^{2}  |\di_{y}\varphi|^{2} \right\}. 
\end{equation}
From (\ref{eq:spectrum}), the primordial power spectrum of the comoving curvature perturbation is hence
\begin{equation}\label{eq:spectrum formula}
	\mathcal{P}_{\mathcal{R}_{c}}(k) = \mathcal{P}_{0}(k) \left[ 2\int_{y_-}^{y_+} dy\, y^{2} |\varphi|^{2}  \right]_{k \ll Ha} \quad \mathcal{P}_{0}(k) = \frac{H^{4}}{4\pi^{2}\dot\phi^{2}}.
\end{equation}
Here, $\mathcal{P}_{0}(k)$ is the standard slow-roll inflation result (as discussed in \cite{Lyth:2009zz}, $\mathcal{P}_{0}(k)$ should be evaluated near horizon crossing).

To summarize, to calculate the primordial power spectrum we need to solve the time-dependent Schr\"odinger equation (\ref{eq:schrodinger 3}) with Bunch-Davies like initial conditions (\ref{eq:Bunch-Davies ICs}), and then calculate the integral in (\ref{eq:spectrum formula}).  Depending on the form of $f$ in the modified commutator (\ref{eq:commutator}),  equation (\ref{eq:schrodinger 3}) may or may not be easy to solve. We now proceed to outline what is analytically known about the solutions of (\ref{eq:schrodinger 3}) for three specific choices of $f$.
 
\subsubsection{Case 1: $f(x)=1$}\label{sec:case 1}

In this case, we have
\begin{equation}
	[ \hat\chi_{\k}, \hat\pi_{\k'}]=i\delta_{\k,\k'}, \quad U(x) = x;
\end{equation}
i.e., we have the standard quantum algebra.  We see that $U(\sqrt{g} y)$ is single valued for $y \in \mathbb{R}$, so we take $y_\pm = \pm\infty$.  Hence 
\begin{equation}
	\mathcal{A} = \tfrac{1}{2} k (y^{2} - \di_{y}^{2}),
\end{equation}
and the Schr\"odinger equation (\ref{eq:schrodinger 3}) becomes exactly that of a simple harmonic oscillator of frequency $k$.  This has the exact (normalized) solution
\begin{equation}\label{eq:SHO exact}
	\varphi(\eta,y) = \sum_{n=0}^{\infty} c_{n} e^{-iE_{n}\eta} \varphi_{n}(y), \quad \sum_{n=0}^{\infty} |c_{n}|^{2} =1,
\end{equation}
where $E_{n}$ and $\varphi_{n}$ are orthonormal solutions of the eigenvalue problem (\ref{eq:eigenvalue problem}), and the $c_{n}$'s are (constant) expansion coefficients.  Explicit formulae for the eigenvalues and eigenfunctions are given in appendix \ref{sec:eigenvalues}.  The Bunch-Davies prescription for initial data (\ref{eq:Bunch-Davies ICs}) implies that we must set 
\begin{equation}\label{eq:case 1 ICs}
	c_{n}(g_{0}) = c_{n}(g)  = \delta_{n,0}.
\end{equation}	
Plugging this into the exact solution and evaluating the integral in (\ref{eq:spectrum formula}) yields 
\begin{equation}
	\mathcal{P}_{\mathcal{R}_{c}}(k) = \mathcal{P}_{0}(k) .
\end{equation}	
That is, for this case we recover the standard slow-roll inflation result for the power spectrum.

\subsubsection{Case 2: $f(x)=1+x^{2}$}

In this case, we have
\begin{equation}
	[ \hat\chi_{\k}, \hat\pi_{\k'}]=i\delta_{\k,\k'} \left( 1 + \frac{\hat\pi_{\k}^{2}}{\M a^{3}} \right), \quad U(x) = \tan x.
\end{equation}
We see that $U(\sqrt{g} y)$ will be single-valued if we take $y_\pm = \pm\pi/2\sqrt{g}$.  The $\mathcal{A}$ operator becomes
\begin{equation}\label{eq:case 2 A}
	\mathcal{A}  = \frac{k}{2} \left[ \frac{\tan^{2}(\sqrt{g}y)}{g} - \frac{\di^{2}}{\di y^{2}} \right] .
\end{equation}
Unlike Case 1 above, we cannot write down a closed form solution of the Schr\"odinger equation (\ref{eq:schrodinger 3}) with this form of $\mathcal{A}$.  However, we can analytically solve the eigenvalue problem (\ref{eq:eigenvalue problem}) assuming Dirichlet boundary conditions.  We obtain an infinite number of solutions for $\varphi_{n}$ and $E_{n}$ labelled by $n = 0, 1, 2 \ldots$ (these are given explicitly in appendix \ref{sec:eigenvalues}).  In this case, the Bunch-Davies initial conditions (\ref{eq:Bunch-Davies ICs}) are explicitly:
\begin{equation}\label{eq:case 2 ICs}
	\varphi(\eta_{0},y) =  \left(\frac{g_{0}}{\pi} \right)^{1/4} \left[ \frac{\Gamma(l_{0}+1)}{\Gamma(l_{0}+1/2)} \right]^{1/2} \cos^{l_{0}}(\sqrt{g_{0}} y), \quad l_0 = \frac{1}{2} + \frac{1}{2} \frac{\sqrt{4+g_{0}^{2}}}{g_{0}},
\end{equation} 
where $g_{0}$ is the value of the DQA coupling at the start of inflation.  In principle, to calculate the primordial power spectrum we could numerically solve the PDE (\ref{eq:schrodinger 3}) with initial conditions (\ref{eq:case 2 ICs}) to obtain $\varphi$ in the $k \ll Ha$ limit.  However, in practice it is preferable to first perform a spectral decomposition of the Schr\"odinger equation in terms of the $\varphi_{n}$ eigenfunctions, solve the resulting ODEs numerically, and hence obtain the primordial power spectrum from (\ref{eq:spectrum formula}).  In Appendix \ref{sec:case 2 numeric}, we describe the details of this procedure.  Our numerical results for the power spectrum are found in \S\ref{sec:results}.

\subsubsection{Case 3: $f(x)=1-x^{2}$}

In this case, we have
\begin{equation}
	[ \hat\chi_{\k}, \hat\pi_{\k'}]=i\delta_{\k,\k'} \left( 1 - \frac{\hat\pi_{\k}^{2}}{\M a^{3}} \right), \quad U(x) = \tanh x.
\end{equation}
We see that $U(\sqrt{g} y)$ is single valued for $y \in \mathbb{R}$, so we take $y_\pm = \pm\infty$.  The $\mathcal{A}$ operator is
\begin{equation}\label{eq:case 3 A}
	\mathcal{A}  = \frac{k}{2} \left[ \frac{\tanh^{2}(\sqrt{g}y)}{g} - \frac{\di^{2}}{\di y^{2}} \right] .
\end{equation}
As in Case 2, the Sch\"odinger equation (\ref{eq:schrodinger 3}) equation does not appear to be solvable analytically, but the associated eigenvalue problem (\ref{eq:eigenvalue problem}) is: the eigenfunctions and eigenvalues are given in appendix \ref{sec:eigenvalues}.  Unlike the previous case, the finite height of the potential implies that we obtain a finite number of normalizable eigenfunctions labelled by
\begin{equation}
	n = 0,1,2 \ldots \text{floor} \left( -\frac{1}{2} + \frac{1}{2} \frac{\sqrt{4+g^{2}}}{g} \right).
\end{equation} 
This implies that the eigenfunctions do not form a basis for $L^{2}(\mathbb{R},dy)$, we cannot solve $i\di_{\eta} \varphi = \mathcal{A}\varphi$ using the spectral methods discussed in appendix \ref{sec:case 2 numeric}.  That is, we are obliged to numerically solve the Schr\"odinger equation directly in order to find $\langle \chi_{\k}^{2} \rangle$.  The appropriate initial data is given by the $n=0$ eigenfunction:
\begin{equation}\label{eq:case 3 ICs}
	\varphi(\eta_{0},y) = \left(\frac{g_0}{\pi} \right)^{1/4} \left[ \frac{\Gamma(j_{0}+1/2)}{\Gamma(j_{0})} \right]^{1/2} \cosh^{-j_{0}}(\sqrt{g_{0}}{y}), \quad	j_0 = -\frac{1}{2} + \frac{1}{2} \frac{\sqrt{4+g_{0}^{2}}}{g_{0}}.
\end{equation}
In Appendix \ref{sec:case 3 numeric}, we describe the numerical methods we use to solve the Schr\"odinger equation in this case.  Our numerical results for the power spectrum are found in \S\ref{sec:results}.

\subsection{Qualitative features and limiting cases}\label{sec:qualitative}

Before we go on to present numerical power spectrum results, it is perhaps useful to analytically determine what kind of general features we should expect to see.  We first notice that in the small coupling $g \ll 1$ limit, the $\mathcal{A}$ operator for cases with modified uncertainty relations (cases 2 and 3) approach the $\mathcal{A}$ operator for the standard unmodified scenario (case 1).  That is, for epochs where the physical wavelength of a mode is larger than the DQA length scale $\sim \M^{-1}$, that mode's evolution will be close to the predictions of ordinary curved space quantum field theory.  Conversely, for epochs when $g \gg 1$ the evolution will be significantly altered.\footnote{Notice that for more general choices of $f$, our definition (\ref{eq:U def}) of the $U$ function appearing in (\ref{eq:schrodinger 3}) implies $U(x) = x + \mathcal{O}(x^{3})$ (c.f.\ equation \ref{eq:U def}).  In turn, this implies that we will always recover case 1 dynamics at late times, irrespective of our choice of $f$.} Furthermore, for a mode with a given comoving wavenumber $k$, the coupling $g = k/\M a$ will be largest at the start of inflation and approach zero near the end.  Hence, the greatest DQA effects will occur at early times.   If the initial coupling is small $g_{0} \ll 1$, it will remain small throughout inflation; which means power spectrum should be close to the standard slow-roll result $\mathcal{P}_{0}(k)$ on larges scales defined by $g_{0} \ll 1$, or equivalently $k \ll \M a_{0}$, where $a_{0}$ is the initial scale factor.  

Next, we note that if we transform the time coordinate in (\ref{eq:schrodinger 3}) from $\eta$ to $g = -\epsilon k\eta$, the only adjustable parameter in the resulting PDE is $\epsilon$.  We also note that the initial conditions [(\ref{eq:case 1 ICs}), (\ref{eq:case 2 ICs}) or (\ref{eq:case 3 ICs})] are completely fixed by selecting $g_{0}$.  These facts mean that for a particular choice of $f$, the quantum evolution of a mode is entirely specified by selecting $\epsilon$ and $g_{0}$; from which it follows that the ratio $\mathcal{P}_{\mathcal{R}_{c}}/\mathcal{P}_{0}$ will be a function of $\epsilon$ and $g_{0}$ only.  

It is obvious that the initial coupling $g_{0}$ plays a central role in determining the essential features of the power spectrum.  It is therefore useful parameterize it in a more physical way:
\begin{equation}
	g_{0} = \frac{k}{k_{\star}}, \quad k_{\star} = \M a_{0} = - \frac{1}{\epsilon\eta_{0}}.
\end{equation}
Here, the DQA pivot scale $k_{\star}$ represents the comoving wavenumber of a mode that has $\lambda_\text{phys} = \lambda_{\star} = 2\pi/\M$ at the beginning of inflation.  The actual numerical value of $k_{\star}$ is straightforward to calculate \cite{Seahra:2012un}:
\begin{align}\nonumber
	k_\star & = \frac{\pi^{1/2}k_\text{b}T_\text{now} \mathcal{G}_\text{now}^{1/3}}{3 \cdot 30^{1/4}} \mathcal{G}_\text{end}^{-1/12} e^{-N} \frac{\M}{H} \frac{E_\text{inf}}{m_\text{Pl}} \\ & \sim \frac{6\times10^{-6}}{\text{Mpc}} \epsilon^{-1}  \left( \frac{E_{\text{inf}}}{10^{16}\,\text{GeV}} \right) \left( \frac{e^{65}}{e^{N}} \right)  \left( \frac{100}{\mathcal{G}} \right)^{1/12}, \label{eq:k star}
\end{align}
Where $T_\text{now} = 3.94\,\text{K}$ is the current temperature of the universe, $\mathcal{G}_\text{now} = 3.04$ is the current number of relativistic species in the universe, $E_{\text{inf}} = \rho_{\inf}^{1/4}$ is the energy scale of inflation,  $N = \ln (a_{\text{end}}/a_{0})$ is the number of $e$-folds of inflation, and $\mathcal{G}$ is the effective number of relativistic species at the end of inflation.

Finally, it is fairly easy to derive the first order deviation of $\mathcal{P}_{\mathcal{R}_{c}}$ from $\mathcal{P}_{0}$ in the large scale $g_{0} \ll 1$ limit (i.e.\ $k \ll k_{\star}$).  Recall that $f$ is a smooth even function with $f(0)=1$, so its Taylor series expansion is
\begin{equation}
	f(x) = 1 + \tfrac{1}{2} f''(0) x^{2} + \cdots
\end{equation}
Then, it follows from (\ref{eq:U def}) that
\begin{equation}\label{eq:small U}
	U(x) = x + \tfrac{1}{6} f''(0) x^{3} + \cdots,
\end{equation}
and we can write the Schr\"odinger equation (\ref{eq:schrodinger 3}) as
\begin{equation}\label{eq:schrodinger 3}
	i \frac{\di \varphi}{\di \eta} = \frac{k}{2} \left[ - \frac{\di^{2}}{\di y^{2}} + y^{2} + \frac{1}{3} f''(0) g y^{4} + \mathcal{O}(g^{2}y^{6}) \right]\varphi.
\end{equation}
Since we are assuming $g_{0} \ll 1$, we have $g\ll 1$ throughout inflation and we can drop the $ \mathcal{O}(g^{2}y^{6})$ terms and treat $ \frac{1}{3} f''(0) g y^{4}$ as a perturbation to the simple harmonic oscillator Hamiltonian.  It is then straightforward to perform a spectral decomposition of this this approximate Schr\"odinger equation analogous to the one presented in appendix \ref{sec:case 2 numeric} and solve the ensuing equations of motion by working consistently to first order in $g_{0}$.  Assuming the Bunch-Davies initial conditions (\ref{eq:Bunch-Davies ICs}), we obtain
\begin{equation}
	\mathcal{P}_{\mathcal{R}_{c}} (k) = \mathcal{P}_{0}(k) \left[ 1 - \frac{1}{2} f''(0) \frac{k}{k_{\star}} + \mathcal{O} \left(\frac{k^{2}}{k_{\star}^{2}}\right) \right] 
\end{equation}
From this it is manifestly obvious that we recover $\mathcal{P}_{\mathcal{R}_{c}} (k) \approx \mathcal{P}_{0}(k)$ on large scales $k \ll k_{\star}$.

\section{Results}\label{sec:results}

\subsection{Case 2: $f(x)=1+x^{2}$}

\begin{figure}
\begin{center}
\includegraphics[scale=0.35]{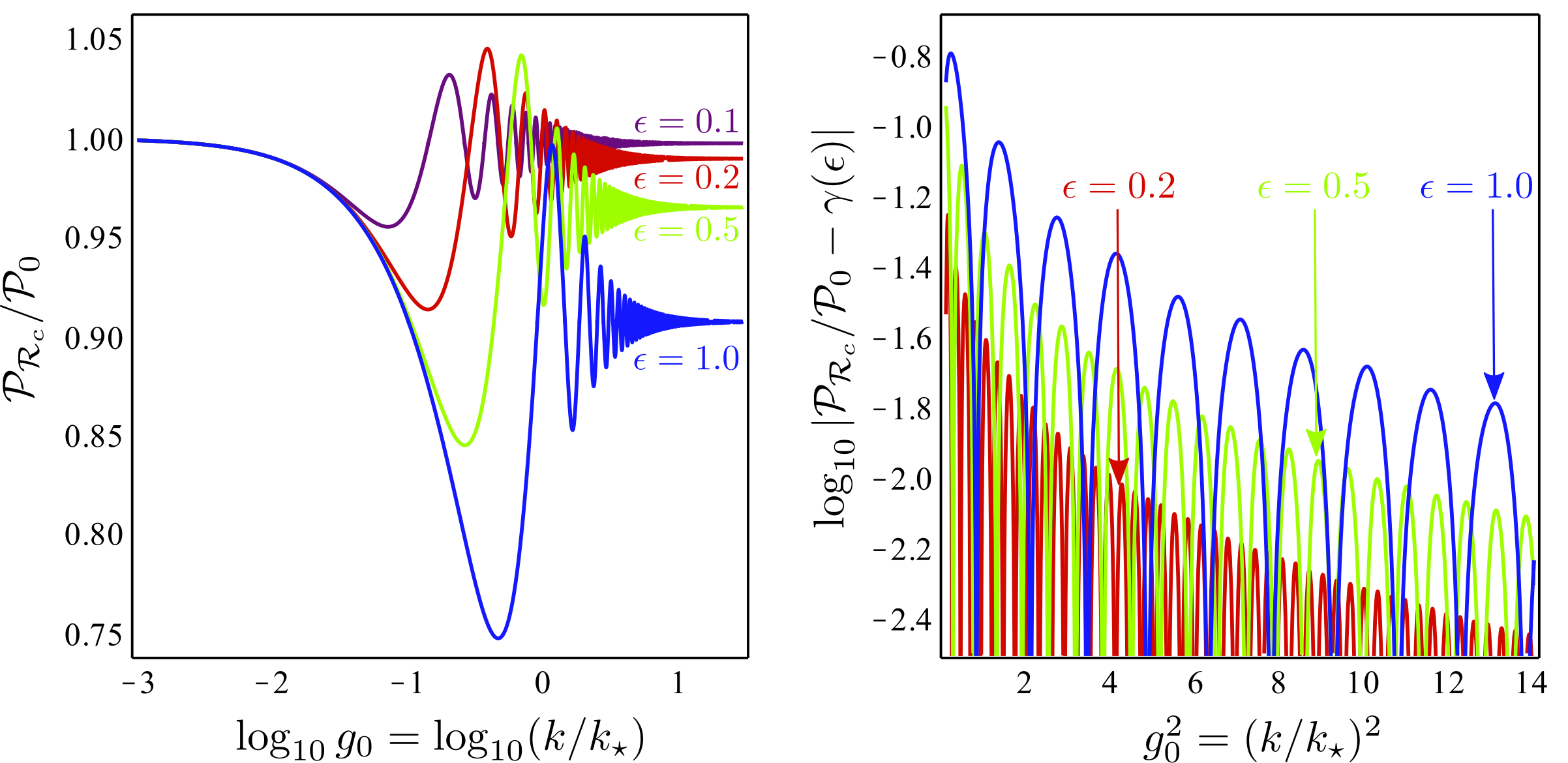}
\end{center}
\caption{Numerical results for the power spectrum for Case 2.  Here, $\mathcal{P}_{0}$ refers to the power spectrum obtained from conventional uncertainty relations (Case 1).  On large scales $k \ll k_{\star}$, we find $\mathcal{P}_{\mathcal{R}_{c}} \approx \mathcal{P}_{0}$ as expected.}\label{fig:case 2 primordial}
\end{figure}
In figure \ref{fig:case 2 primordial}, we plot our numerical results for the case 2 primordial power spectrum for several values of $\epsilon = H/\M$.  The key features of the spectra are as follows:
\begin{itemize}
	\item For small $g_{0} = k/k_{\star}$, we see that the modified spectra approach the slow-roll result, as expected.
	\item For large $g_{0} = k/k_{\star}$, the spectra approach a $\epsilon$-dependant constant, which we call $\gamma(\epsilon)$.  It is apparent that $\gamma(\epsilon) \rightarrow 1$ as $\epsilon \rightarrow 0$.  The right panel of figure \ref{fig:case 2 primordial} shows the deviations of the spectra from $\gamma(\epsilon)$, which take the form of decaying oscillations evenly spaced in $g^{2}$. 
	\item The largest deviations from the slow-roll result occur at $g_{0} = k/k_{\star} \sim 1$.  The magnitudes of the deviation decrease with decreasing $\epsilon$. 
\end{itemize}
From examining numeric simulations with $\epsilon$ between $0.01$ and $1$, we find the following fits for our spectra on large and small scales
\begin{equation}
	\frac{\mathcal{P}_{\mathcal{R}_{c}}}{\mathcal{P}_{0}} \approx \begin{cases}
	1 - g_{0}, & g_{0} \ll 1, \\ 1 -0.215\,\epsilon^{2}+0.125\,\epsilon^{3} + 0.215\,\epsilon g_{0}^{-2} \sin(2\epsilon^{-1}g_{0}^{2}), & g_{0} \gg 1.
	\end{cases}
\end{equation}
The CMB angular power spectrum is found from an integration of the power spectrum across the brightness functions $\Delta(k,\eta_0)$, which are highly oscillating functions characterising the fluid evolution between the end of inflation and the surface of last scattering:
\begin{equation}
C_\ell\propto \int\mathcal{P}_{\mathcal{R}_{c}}(k)\left|\Delta(k,\eta_0)\right|^2d(\ln k) .
\end{equation}
In what follows, we will parametrize the standard slow-roll spectrum by
\begin{equation}
\mathcal{P}_0(k)=A\left(\frac{k}{k_0}\right)^{n_s-1}
\end{equation}
for some pivot scale $k_0$.

In this case, the power spectrum is modulated on smaller scales by rapidly damping oscillations, in general of a different phase to those in the brightness oscillations. This can therefore be expected to cause complex deviations, particularly at higher $\ell$-values where smaller scales have more imprint.  Further, $\mathcal{P}_{\mathcal{R}_c}$ is equal to $\mathcal{P}_0$ for large scales, larger than $\mathcal{P}_0$ across a narrow band, and tends towards a constant suppression of $\mathcal{P}_0$ on small scales. With no rescaling of the primordial amplitude we therefore expect to see a small boost in the CMB angular power spectrum across a narrow region and a suppression in the damping tail for high $\ell$. In practice we can normalize the amplitude arbitrarily and typically choose to either normalize at the first acoustic peak (at $\ell=226$) or at high $\ell$ (at $\ell=2000)$.

\begin{figure}
\begin{center}
\includegraphics[width=\textwidth]{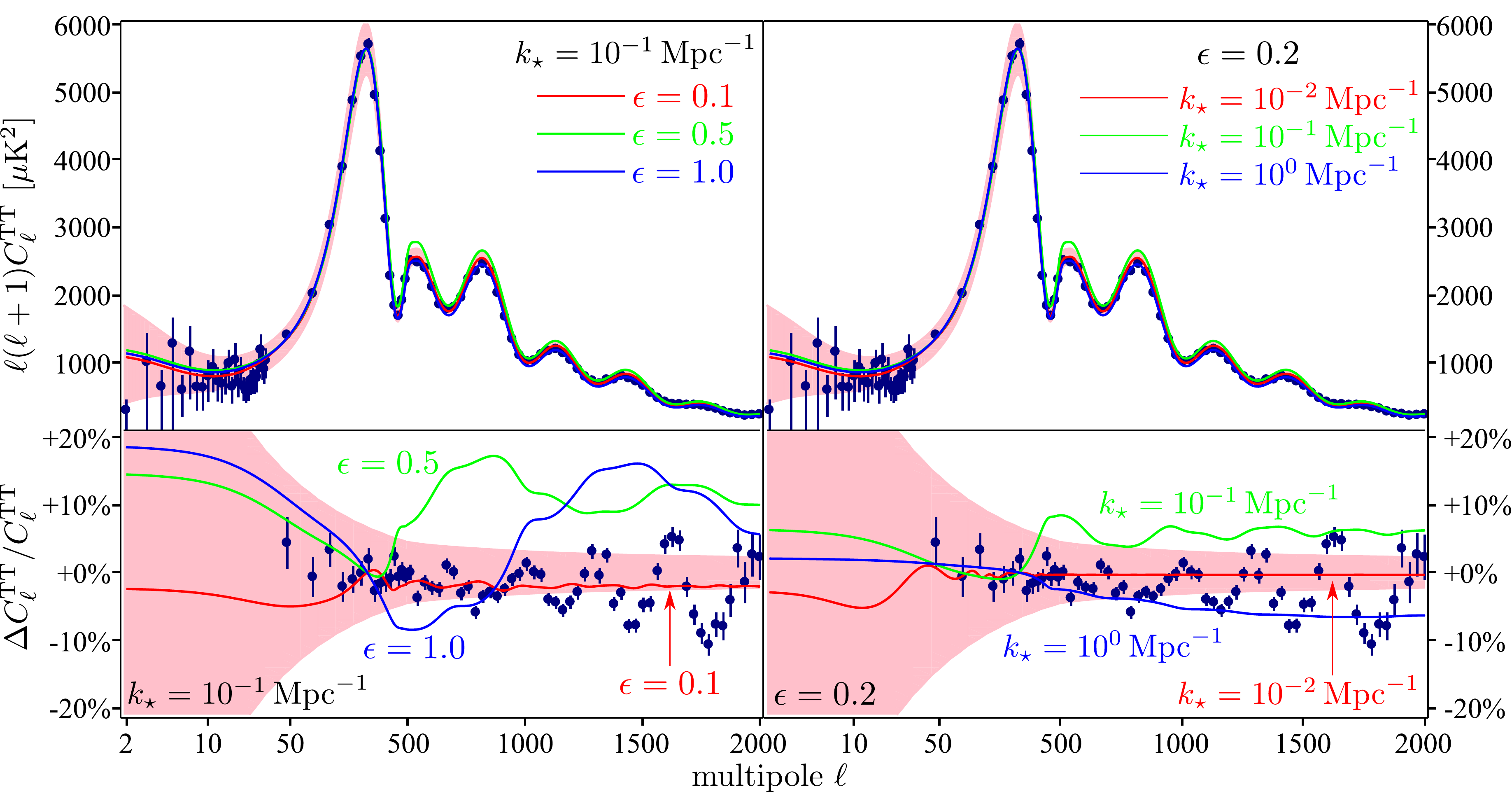}
\caption{CMB temperature-temperature (TT) power spectra generated from case 2 primordial power spectra for various choices of $k_{\star}$ and $\epsilon$ (top panels).  We have assumed cosmological parameters from the best fit to the 2013 Planck + WMAP polarization data release.  We take the unmodified case 1 power spectrum to be $\mathcal{P}_{0} = A (k/k_0)^{n_{s}-1}$ with $n_{s} = 0.9603$.  We have selected $A$ such that all case 2 TT spectra match predictions from unmodified uncertainty relations (case 1) at the position of the first acoustic peak ($\ell = 226$).   We also show relative deviations of the modified spectra from the unmodified case 1 results (bottom panels).  The dark blue points and error bars are the Planck results \cite{Planck:2013kta}, which are binned for $\ell > 31$ (noisy $\ell \le 31$ data has been omitted on the bottom panels for clarity).  On all plots, the pink shaded areas indicate uncertainties due to cosmic variance in the best-fit unmodified spectra.}\label{fig:case 2 CMB}
\end{center}
\end{figure}

In figure \ref{fig:case 2 CMB}, we plot the CMB temperature-temperature power spectra obtained from the primordial spectra of figure \ref{fig:case 2 primordial} using brightness functions recovered from the CAMB software package \cite{Lewis:1999bs} and wrapped across the distorted spectrum with a high-precision integrator capable of sampling the oscillations.  We employ the best fit Planck 2013 + WMAP Polarization $\Lambda$CDM cosmology found in the Planck 2013 data release \cite{Ade:2013zuv}. In particular, the Hubble rate is $h=0.6704$, baryon and CDM density parameters $\Omega_bh^2=0.022032$ and $\Omega_ch^2=0.12038$ and the universe is taken to be flat. The amplitude of the primordial perturbations for a standard uncertainty relation is $A=2.2154\times 10^{-9}$ at $k_0=0.05 \,h/\text{Mpc}$ and the spectral index is $n_s=0.9603$. The amplitudes of the case 2 spectra have been adjusted such that they match the unmodified case at the first peak, $\ell=226$.

Also shown in figure \ref{fig:case 2 CMB} are deviations of the case 2 CMB spectra from the the best-fit model with standard uncertainty relations (case 1), and the cosmic variance uncertainty band about the best-fit model.  As anticipated, the results show a complicated mixing of oscillations, tending towards a constant shift for very low and very high $\ell$.

\begin{figure}
\begin{center}
\includegraphics[width=0.8\textwidth]{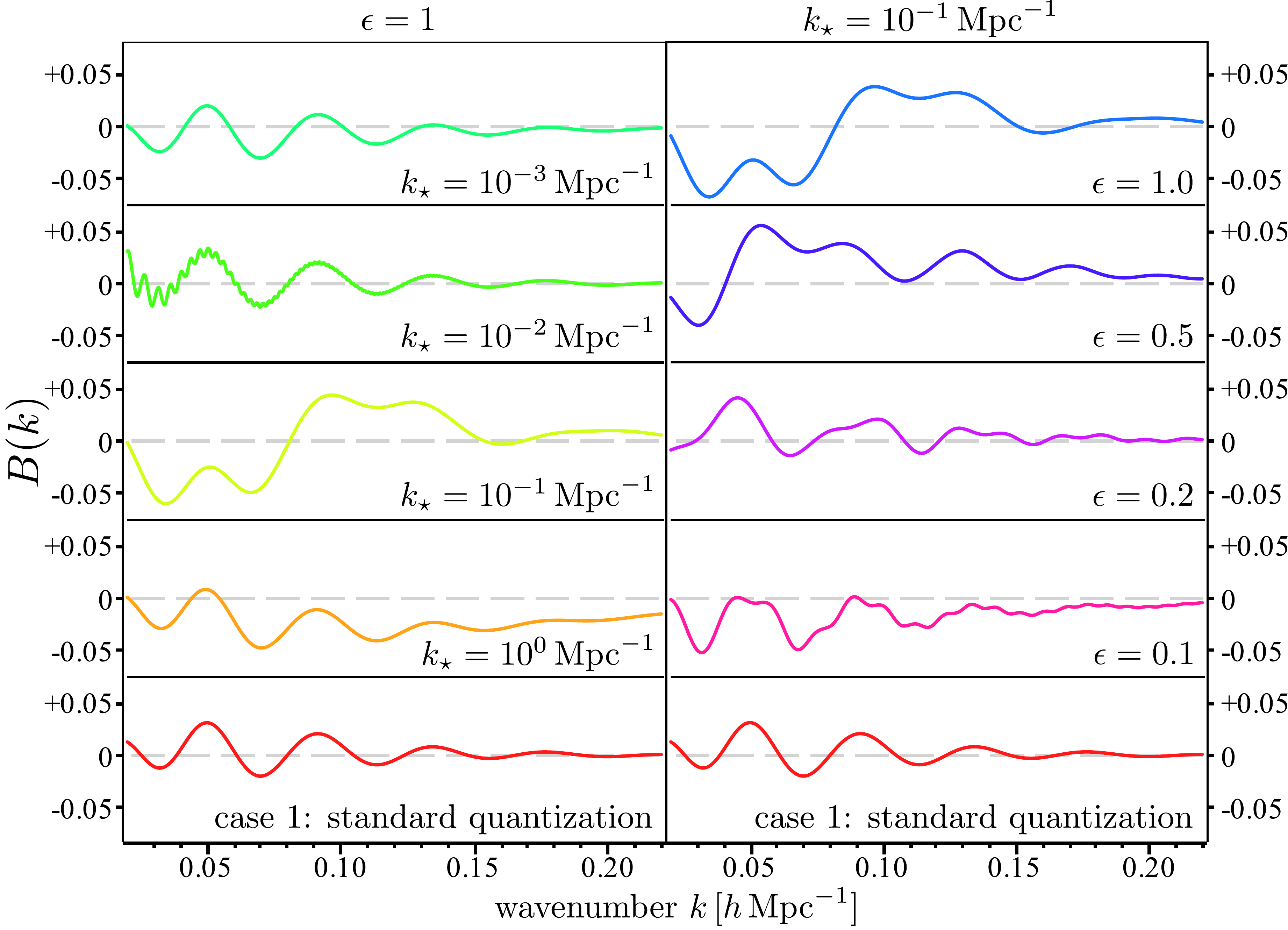}
\end{center}
\caption{Baryon acoustic oscillations for case 2}\label{fig: case 2 BAO}
\end{figure}

The other main dataset that is of increasing importance to modern cosmology is the distribution of large-scale structure, which is under ongoing study by the Sloan Digital Sky Survey (SDSS) \cite{Eisenstein:2011sa} and WiggleZ \cite{Parkinson:2012vd} projects, along with upcoming projects such as Euclid \cite{Laureijs:2011gra}. The key observables for our purposes are the matter power spectrum
\begin{equation}
P(k)=\frac{2\pi^2}{k^3}\mathcal{P}_{\mathcal{R}_c(k)}\left|\delta_m(k)\right|^2
\end{equation}
where $\delta=\delta\rho/\overline{\rho}$ is the matter overdensity and, in particular, the baryon acoustic oscillations (BAOs) observed (for instance) by \cite{Eisenstein:2005su,Cole:2005sx,Anderson:2013oza,Blake:2011en} which provide an extremely sensitive probe of the composition and evolution of the universe. Recovering the BAOs from a matter power spectrum involves determining a smoothed power spectrum without oscillations; the SDSS collaboration have employed various techniques to recover this, such as taking splines centred around carefully chosen nodes, or employing the oscillation-free fitting formulae of \cite{Eisenstein:1997ik}. We choose to recover the smoothed spectra in a model-independent manner employing a running average on the logarithm of the matter power spectrum.\footnote{One advantage of this approach is that it does not induce a spuriously large signal on large scales---the recovered BAOs decay rapidly as k approaches the homogeneity scale, as is obvious from the power spectrum \cite{Anderson:2013oza}.  A corresponding disadvantage is that the smoothed spectrum is sensitive to the window function of the running average; we took a top-hat and adjusted its width in $k$.} From the smoothed spectrum $P_\text{smooth}(k)$ the baryon acoustic oscillations can be characterised \cite{Anderson:2013oza} by
\begin{equation}
B(k)=
\left[\frac{P(k)}{P_\text{smooth}(k)}-1\right]\exp\left(-k^2\Sigma_{\rm NL}^2\right)+1.
\end{equation}
Here $\exp(-k^2\Sigma_{\rm NL}^2)$ is a non-linear smoothing employed to model the damping of the BAOs by non-linear processes, and we choose $\Sigma_{\rm NL}=8.24\, \text{Mpc}/h$, employed by \cite{Anderson:2013oza} for their pre-reconstruction fits. The results are shown in figure \ref{fig: case 2 BAO}, with all oscillation plots taken relative to the smoothed $P(k)$ from the standard case. This not least enables us to compare the figures with the data, which are present relative to the standard model.  Qualitatively, the results are again as should be broadly expected: for combinations of $\epsilon$ and $k_\star$ for which the scale of the oscillations in $\mathcal{P}_{\mathcal{R}_c}(k)$ coincide with the BAOs we recover large deviations from the standard BAO prediction, and the deviations die off on both very large and very small scales.

For this class of modified uncertainty relations, it is apparent that it is not difficult to find choices of $\epsilon$ and $k_{\star}$ that generate CMB and BAO predictions whose  discrepancy with the standard model is within an acceptable range.  The reason is that the primordial spectra for $k \gtrsim k_\star$ matches the unmodified result $\mathcal{P}_{0}$ with a overall constant reduction in amplitude; i.e., $\mathcal{P}_{\mathcal{R}_{c}}$ has the same shape as $\mathcal{P}_{0}$ on small scales.   However, the question of whether or not any case 2 spectra give a \emph{better} fit to the data than the standard uncertainty relation case is much more involved and beyond the scope of this paper.

\subsection{Case 3: $f(x)=1-x^{2}$}

\begin{figure}
\includegraphics[scale=0.25]{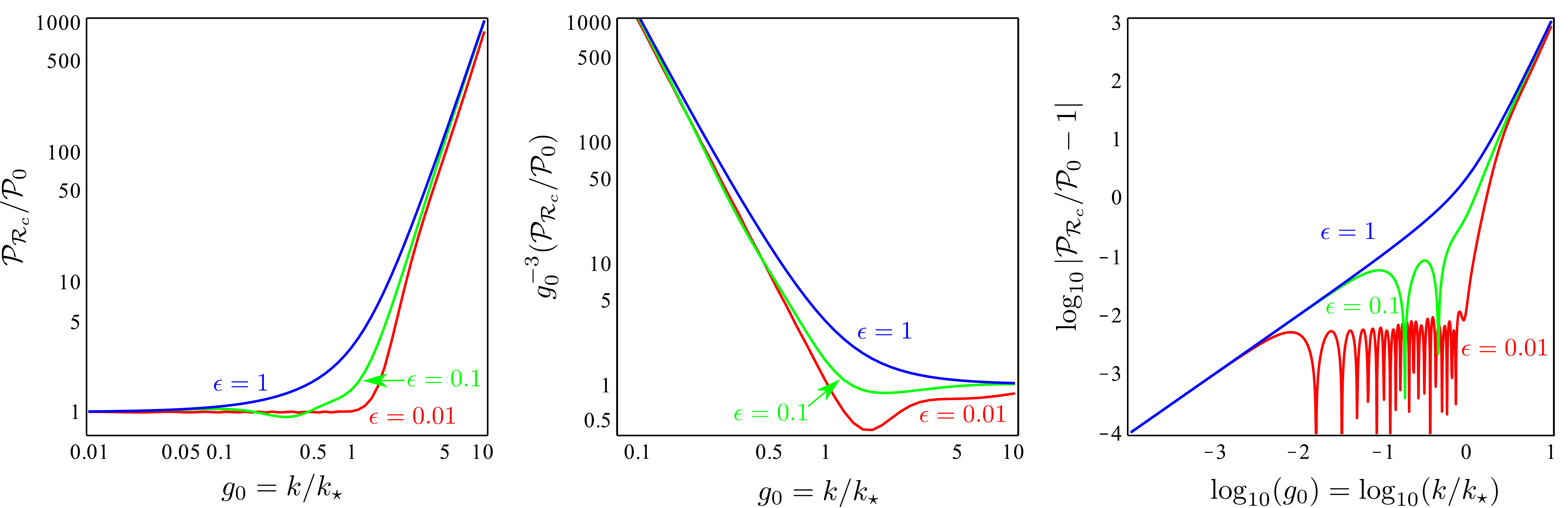}
\caption{Numerical results for the power spectrum for Case 3.  Here, $\mathcal{P}_{0}$ refers to the power spectrum obtained from conventional uncertainty relations (Case 1).  On large scales $k \ll k_{\star}$, we find $\mathcal{P}_{\mathcal{R}_{c}} \approx \mathcal{P}_{0}$ as expected.  However, on small scales $k \gtrsim k_{\star}$ we see a very blue spectrum with $\mathcal{P}_{\mathcal{R}_{c}}/\mathcal{P}_{0} \propto k^{3}$, which represents a large deviation from Schr\"odinger quantization.  The relative deviation between the case III and case I power spectra is plotted in the righthand panel, where we see that spectra from the modified uncertainty relation is oscillatory on intermediate scales $k \sim k_{\star}$ when $\epsilon = H/M_{\star} \lesssim 1$.}\label{fig:case 3 primordial}
\end{figure}

In figure \ref{fig:case 3 primordial}, we plot our numerical results for the case 3 primordial power spectrum for several values of $\epsilon = H/\M$.  The key features of the spectra are as follows:
\begin{itemize}
	\item For small $g_{0} = k/k_{\star}$, we see that the modified spectra approach the slow-roll result, as expected.
	\item For large $g_{0} = k/k_{\star}$ we see that the spectrum diverges like $k^{3}$; i.e.,  $\mathcal{P}_{\mathcal{R}_{c}} \approx g_{0}^{3} \mathcal{P}_{0}$, irrespective of the value of $\epsilon$. 
	\item For intermediate, $g_{0} = k/k_{\star} \sim 1$, the differences between the computed and slow-roll power spectra depend on $\epsilon$, and we see oscillations that become more numerous as $\epsilon \rightarrow 0$.  
\end{itemize}
Clearly, the most striking feature of the spectra is the $k^{3}$ ultraviolet divergence, which represents a very large modification of the slow-roll result on small scales.  This kind of radical departure may have been anticipates from the form of the $\mathcal{A}$ operator (\ref{eq:case 3 A}) in case 3:  For large $g$, this operator supports only a single normalizable eigenfunction, whereas the dimension of the eigenbasis is infinite in cases 1 and 2 for all $g$.  That is, the operator for case 3 is qualitatively much different from the other cases.  The fact that $\mathcal{P}_{\mathcal{R}_{c}}/\mathcal{P}_{0} \propto k^{3}$ at high $k$ is a bit puzzling, but we can speculate it has something to do with the initial data:  If we evaluate the $\langle \chi_{\k}^{2} \rangle$ integral (\ref{eq:chi squared}) at the start of inflation using the initial data (\ref{eq:case 3 ICs}), we find $\langle \chi_{\k}^{2} \rangle_{0} \approx \frac{1}{2} H^{2} k^{-3} g_{0}^{3}$ in the $g_{0} \rightarrow \infty$ limit.  This means that if we were to calculate the fluctuation spectrum at the beginning of inflation (as opposed to the end) we would find  $\mathcal{P}_{\mathcal{R}_{c}}/\mathcal{P}_{0} \approx k^{3}/k_{\star}^{3}$.  Stated another way, the UV divergence of the final spectrum of fluctuations seems to be a property inherited from the Bunch-Davies like initial conditions we have enforced.

\begin{figure}
\includegraphics[width=\textwidth]{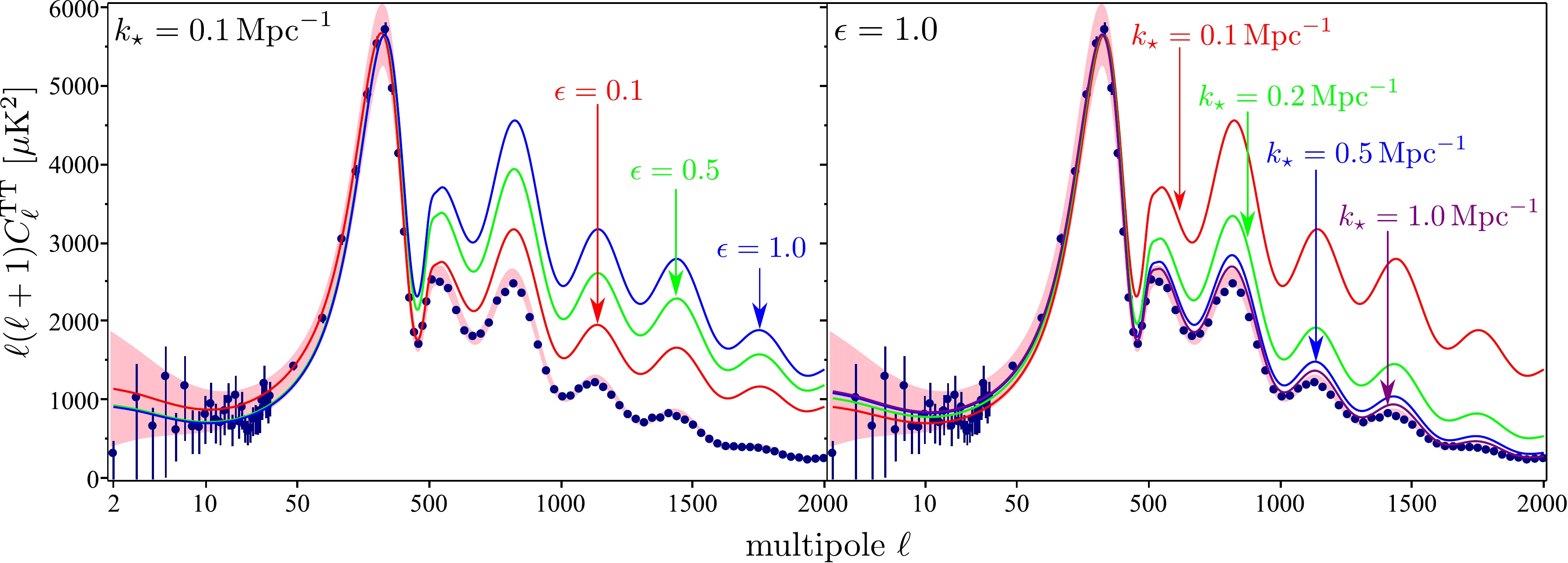}
\caption{CMB temperature-temperature (TT) power spectra generated from case 3 primordial power spectra.  As in figure \ref{fig:case 2 CMB}, we have assumed cosmological parameters from the best fit to the 2013 Planck data release.  The dark blue points and error bars are the Planck results.  We take the unmodified power spectrum to be $\mathcal{P}_{0} = A k^{n_{s}-1}$ with $n_{s} = 0.9603$.  We have selected $A$ such that all case 2 TT spectra match predictions from unmodified uncertainty relations (case 1) at the position of the first acoustic peak ($\ell \sim 120$).  The pink shaded areas indicate uncertainties due to cosmic variance in the best-fit unmodified spectra.}\label{fig:case 3 CMB}
\end{figure}

Somewhat predictably, the extreme features of the primordial spectra in this case lead to poor agreement with CMB observations.  In figure \ref{fig:case 3 CMB}, we plot the CMB TT spectra obtained for this case using the same assumptions as for figure \ref{fig:case 2 CMB}.   The only way to obtain a reasonable agreement with Planck data is to set $k_{\star} \gtrsim 1.0 \, \text{Mpc}^{-1}$, which has the effect of pushing the UV divergences to scales irrelevant for the CMB.  Equation (\ref{eq:k star}) implies that such a large value of $k_{\star}$ involves small $\epsilon$, high energy inflation $E_\text{inf} \gg 10^{16}\,\text{GeV}$, short inflation $N < 65$, or some combination of these effects.

It is impossible to reconcile the case 3 spectrum with the CMB without effectively reducing it to the standard unmodified case, or else selecting parameters inconsistent with slow-roll. As a result it is not necessary to present the BAO results; since the primordial spectrum is directly mapped onto the matter power spectrum it is obvious that the predictions will deviate vastly from observation.\footnote{Furthermore, the $(k/k_\star)^3$ divergence would prove extremely problematic for studies of structure formation on cluster and galactic scales.}

\section{Discussion}\label{sec:discussion}

In this paper, we have considered the possibility that the algebra of quantum operators gets modified at high energies and calculated the consequences of such a modification on the generation of primordial density fluctuations.  

More specifically, we have developed a general formalism to calculate the perturbative power spectrum when the commutator between the Fourier scalar field amplitude $\chi_{\k}$ and momentum $\pi_{\k}$ gets modified to $[ \hat\chi_{\k}, \hat\pi_{\k'}]=i\delta_{\k,\k'} f ( \hat\pi_{\k}\M^{-1/2}a^{-3/2} ),$ where $f$ is a smooth even function satisfying $f(0)=1$, and $\M$ is a mass scale.  We showed that in order to calculate the spectrum, one needs to solve a Schr\"odinger equation with a time-dependant potential and initial conditions that correspond to a Bunch-Davies like vacuum.  

For the specific cases $f(x) = 1 \pm x^{2}$, we solved for the spectrum numerically.  In the case $f(x) = 1+x^{2}$, we found that the spectrum received small scale corrections in the form of an overall offset of $\mathcal{O}(H^{2}/\M^{2})$ with superimposed oscillations that damp to zero for $k \rightarrow \infty$.  For the case $f(x) = 1 -x^{2}$, we found that the spectrum exhibited a strong ultraviolet divergence $\propto k^{3}$.  We obtained predictions for the cosmic microwave background radiation and baryon acoustic oscillations using our numerically obtained spectra.  For the $f(x) = 1 + x^{2}$, it was not difficult to find scenarios that resulted in predictions close to observations, while the only way to reconcile the $f(x) = 1-x^{2}$ case was to resort to extreme parameter choices.

The obvious next step in this programme should be the detailed comparison of the $f(x) = 1+x^{2}$ predictions to Planck and other observations.  The Planck collaboration has already compared some models with oscillatory spectra to their results, but none of those scenarios involved a spectrum that decayed to a $\mathcal{O}(H^{2}/\M^{2})$ constant offset at high-$k$ \cite{Ade:2013uln}.  This will be the focus of future work.

\appendix

\section{Eigenvalue problem solutions}\label{sec:eigenvalues}

Consider the eigenvalue problem
\begin{equation}
	E_{n} \varphi_{n} = \mathcal{A} \varphi_{n},
\end{equation}
where the differential operator $\mathcal{A}$ is given by
\begin{equation}
\mathcal{A} = \frac{k}{2} \times \begin{cases}

y^{2} - \di_{y}^{2}, & \text{case 1}, \\
 
g^{-1} \tan^{2}(\sqrt{g}y) -  \di_{y}^{2} , & \text{case 2}, \\
 
g^{-1} \tanh^{2}(\sqrt{g}y)-  \di_{y}^{2}, & \text{case 3}.

\end{cases}
\end{equation}
We take $y \in (y_{-},y_{+})$ with
\begin{equation}
y_{\pm} =  \pm \begin{cases}

\infty, & \text{case 1}, \\
 
\tfrac{1}{2} \pi g^{-1/2} , & \text{case 2}, \\
 
\infty, & \text{case 3}.

\end{cases}
\end{equation}
Then, orthonormal solutions $\langle \varphi_{n} | \varphi_{m} \rangle = \delta_{n,m}$ (c.f.\ equation \ref{eq:inner product}) to the eigenvalue problem satisfying Dirichlet boundary conditions are:
\begin{equation}
\varphi_{n} = N \times \begin{cases}

e^{-y^{2}/2 } H_{n}(y), & \text{case 1}, \\
 
 [1+ \tan^{2}(\sqrt{g}y) ]^{-(n+l)/2} C_{n}^{(-n-l+\frac{1}{2})}\left( i\tan \sqrt{g}y \right) , & \text{case 2}, \\
 
 [1- \tanh^{2}(\sqrt{g}y) ]^{-(n-j)/2} C_{n}^{(-n+j+1/2)}\left(\tanh \sqrt{g}y \right) , & \text{case 3};

\end{cases}
\end{equation}
with eigenvalues:
\begin{equation}
E_{n} = \begin{cases}

k(n+1/2), & \text{case 1}, \\
 
 \tfrac{1}{2} k g (n^{2} + 2nl + l) , & \text{case 2}, \\
 
\tfrac{1}{2} kg (-n^{2} + 2nj + j), & \text{case 3};

\end{cases}
\end{equation}
and normalization coefficients:
\begin{equation}
N = \begin{cases}

(\pi^{1/4}\sqrt{2^{n}n!})^{-1}, & \text{case 1}, \\
 
\pi^{-1} g^{1/4} e^{-in\pi/2} 2^{-n-l+\frac{1}{2}} \cos(\pi l) \Gamma(-n-l+\tfrac{1}{2}) \sqrt{(n+l) \Gamma(n+1) \Gamma(n+2l)} , & \text{case 2}, \\
 
\pi^{-1} g^{1/4} e^{i(n+1)\pi/2} 2^{-n+j} \Gamma(-n+j+\tfrac{1}{2}) \sqrt{(j-n) \sin(2\pi j) \Gamma(n+1) \Gamma(n-2j)}, & \text{case 3}.

\end{cases}
\end{equation}
The integer $n$ takes on values
\begin{equation}
n = \begin{cases}

0,1,2 \ldots, & \text{case 1}, \\
 
0,1,2 \ldots, & \text{case 2}, \\
 
0,1,2 \ldots \text{floor}(j), & \text{case 3}.

\end{cases}
\end{equation}
In these formulae, $H_{n}(y)$ and $C_{n}^{(\sigma)}(x)$ are the Hermite and  ultraspherical (Gegenbauer) polynomials of order  $n = 0, 1, 2 \ldots$, respectively \cite{abramowitz+stegun}.  Furthermore, the quantities $l > 1$ and $j>0$ are defined by
\begin{equation}
	l = \frac{1}{2} + \frac{1}{2} \frac{\sqrt{4+g^{2}}}{g}, \quad j = -\frac{1}{2} + \frac{1}{2} \frac{\sqrt{4+g^{2}}}{g}, \quad g^{2} = \frac{1}{l(l-1)} = \frac{1}{j(j+1)}.
\end{equation}
All the eigenfunctions given in this appendix have ever parity for $n$ even and odd parity for $n$ odd.

\section{Numerical methods}\label{sec:numeric}

\subsection{Case 2}\label{sec:case 2 numeric}

To numerically obtain the power spectrum in this case, we perform a spectral decomposition of the Schr\"odinger equation (\ref{eq:schrodinger 3}) in terms of the case 2 eigenfunctions $\varphi_{n}$ listed in appendix \ref{sec:eigenvalues}.  The fact that $\{\varphi_{n}\}$ forms a complete basis of $L^{2}([y_-,y_+],dy)$ allows us to write any solution of (\ref{eq:schrodinger 3}) as
\begin{equation}\label{eq:wavefunction decomposition}
	\varphi(\eta,y) =\sum_{n=0}^{\infty} c_{n}(\eta) e^{i\theta_{n}(\eta)} \varphi_n(\eta,y),  \quad \frac{d\theta_{n}}{d\eta} = -E_{n}.
\end{equation}
It is convenient at this stage to switch from the conformal time coordinate to the dimensionless coupling $g = -k\eta\epsilon$. Some standard quantum mechanical calculations reveal that (\ref{eq:wavefunction decomposition}) will be a solution of (\ref{eq:schrodinger 3}) with (\ref{eq:case 2 A}) if the time dependent expansion coefficients satisfy
\begin{equation}\label{eq:Case 2 ODEs}
	\frac{dc_{n}}{dg} = \sum_{m=0}^{\infty} a_{nm} c_{m},
\end{equation}
where
\begin{equation}
	a_{nm} = \begin{cases} 0,& n=m, \\ \displaystyle -\frac{k \alpha_{mn} \exp(i\epsilon^{-1}\Theta_{mn})}{2(E_{m} - E_{n})} ,  & n\ne m, \end{cases}
\end{equation}
with
\begin{equation}
	\Theta_{mn} = \int_{0}^{g} d\tilde{g} \, \left[ \frac{E_{m}(\tilde{g})-E_{n}(\tilde{g})}{k} \right], \quad \alpha_{mn} = \int_{y_-}^{y_+} dy \, \varphi_{m} \frac{d}{dg} \left[ \frac{\tan^{2}(\sqrt{g}y)}{g} \right] \varphi_n.
\end{equation}
Notice that given the Case 2 eigenfunctions listed in appendix \ref{sec:eigenvalues}, it is possible to calculate the $\alpha_{mn}$ matrix elements analytically. Also, note that from symmetry $\alpha_{nm}$ will only be nonzero if the product $\varphi_{m}\varphi_{n}$ is even; i.e., $\alpha_{mn} = 0$ if $m+n$ is odd.   In this approach, the initial condition (\ref{eq:case 2 ICs}) becomes
\begin{equation}
	c_{n}(g_{0}) = \delta_{n,0}.
\end{equation}
To obtain a numerical solution, we truncate (\ref{eq:Case 2 ODEs}) as follows:
\begin{equation}\label{eq:truncated ode sys}
	\frac{dc_{n}}{dg} = \sum_{m=0}^{M-1} a_{nm} c_{m}.
\end{equation}
Here, the integer $M$ is a cutoff representing the finite dimension of the truncated system.  Because $a_{nm} = 0$ if $n+m$ is odd, the initial data implies that $c_{n} \equiv 0$ for odd values of $n$.  

Before describing our numerical method to solve this system, it is useful to examine a perturbative solution of (\ref{eq:truncated ode sys}) valid for large $g$.  In this regime, we expand about the instantaneous ground state as follows:
\begin{equation}
	c_{n} = \delta_{n,0} + \delta c_{n}, \quad |\delta c_{n}| \ll 1 , \quad \delta c_{n}(g_{0}) = 0.
\end{equation}
Plugging this into (\ref{eq:truncated ode sys}), expanding the amplitude and phase of $a_{nm}$ to leading order in $1/g$, and solving yields:
\begin{equation}
	\delta c_{n} =  \frac{n+1}{2n(n+2)} \left\{  \text{Ei}\left[ i,  \frac{in(n+2)g_{0}^{2}}{4\epsilon} \right] - \text{Ei}\left[ i,  \frac{in(n+2)g^{2}}{4\epsilon} \right] \right\}, \quad n = 2,4,6 \ldots,
\end{equation}
where $\text{Ei}(a,z)$ is the exponential integral and all other $\delta c_{n}$'s are zero.  We see that the perturbative solution oscillates as $g^{2}$ for large $g$; i.e., it undergoes very rapid oscillations at early times.  This suggests that $g$ is not a very good time coordinate for numerical simulations.  Hence, we consider a different time parameter $\tau$ that satisfies $\tau \propto g^{2}$ at large $g$.  For late times, it makes sense to take $\tau \propto \ln g$ (which is equivalent to $\tau \propto Ht$) in order to sufficiently resolve the superhorizon evolution of the mode.  This leads to our choice
\begin{equation}
	g = \sqrt{\text{W}(e^{\tau})}, \quad \tau = 2 \ln g + g^{2},
\end{equation}
where $\text{W}$ is the Lambert-W function.

After transforming the time coordinate, the system can be recast as a single matrix ODE for the expansion coefficients
\begin{equation}\label{eq:matrix ODE}
	\frac{d \vec{c}}{d\tau} = A \vec{c}, \quad \vec{c} = [c_{0} \cdots c_{M-1}]^{T}, \quad A = \frac{dg}{d\tau} [a_{nm}].
\end{equation}
Note that $A=A(\tau)$ is a symmetric, anti-Hermitian and tridiagonal $M \times M$ matrix.  We solve the matrix ODE (\ref{eq:matrix ODE}) using a temporal lattice:
\begin{equation}
	\tau_{i} = \tau_{0} + si, \quad \vec{{c}}_{i} = \vec{c}(\tau_{i}),
\end{equation}
where $\tau_{0}$ is an initial time and $s$ is the timestep.  Our temporal stencil is
\begin{equation}\label{eq:stencil 2}
	\vec{c}_{i+1} = U(\tau_{i}) \vec{c}_{i}, \quad U(\tau_{i}) = \left[I-\tfrac{1}{2} s A(\tau_{i}) \right]^{-1} \left[I+ \tfrac{1}{2} s A(\tau_{i}) \right].
\end{equation}
By construction, the evolution matrix is unitary $U^{\dag} U = I$, which means the norm of $\vec\phi$ is conserved: $\vec{c}_{i}^{\dag} \vec{c}_{i}= \vec{c}_{i+1}^{\dag} \vec{c}_{i+1}$; i.e., our numerical stencil explicitly conserves the normalization of the wavefunction.  Repeated application of (\ref{eq:stencil 2}) allows us to evaluate the expansion coefficients in the $k \ll Ha$ limit.  We then re-construct the wavefunction using a truncated version of (\ref{eq:wavefunction decomposition}), and hence obtain the power spectrum using (\ref{eq:spectrum formula}).

\begin{figure}
\begin{center}
\includegraphics[width=\textwidth]{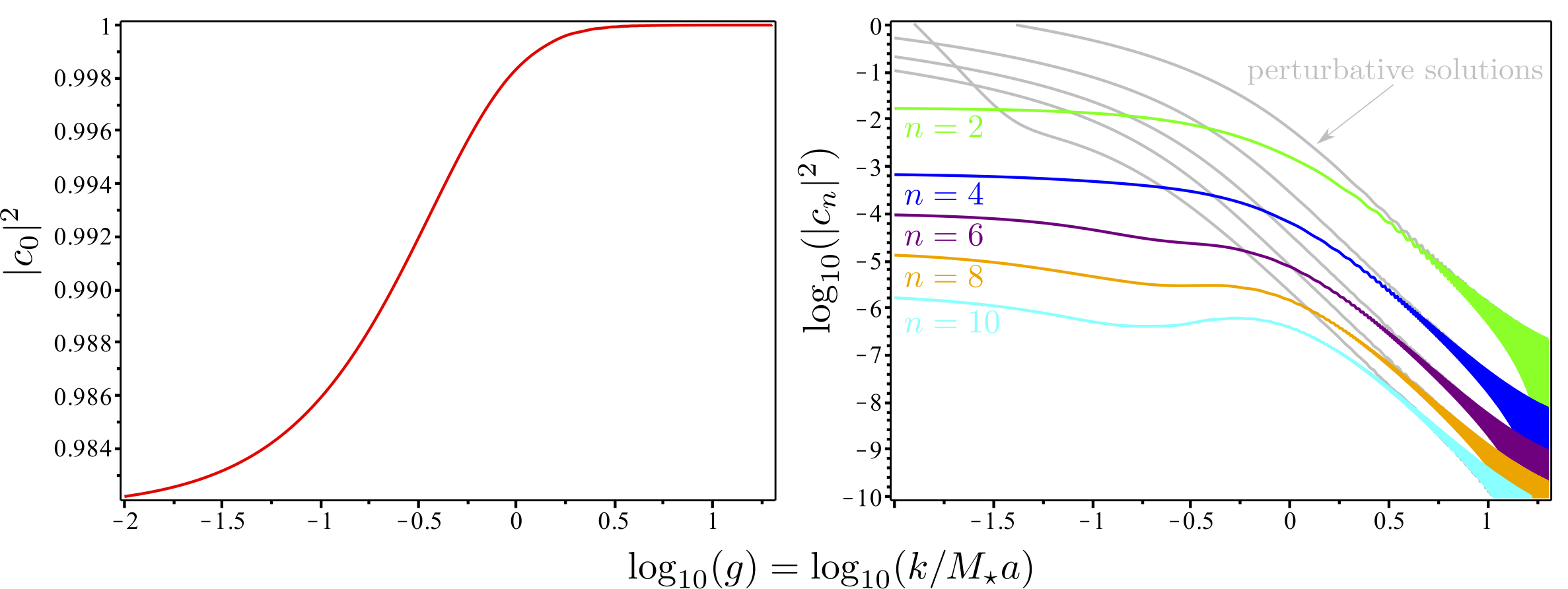}
\end{center}
\caption{An example of our numerical results for the evolution of the probability amplitudes $|c_{n}|^{2}$ in Case 2.  For this simulation, we selected $M=11$, $\epsilon = 1$ and $g_{0} = 20$.  Note that our choice of initial data implies $c_{n} \equiv 0$ for $n$ odd.  Large $g$ perturbative solutions for the amplitudes are shown in grey.}\label{fig:c plot}
\end{figure}

For the results presented in this paper, we truncated the system with cutoff $M = 11$.  An example of the numerical solutions we obtain for the expansion coefficients is given in Figure \ref{fig:c plot}.  This plot illustrates that our numerical results closely match the perturbative solutions for large $g$ (i.e.\ early times).  Also, we see that the magnitudes of the expansion coefficients decreases rapidly with increasing $n$ with $|c_{10}(g)| \ll |c_{0}(g)|$; similar behaviour is observed for other choices of $\epsilon$ and $g_{0}$.  This suggests to us that our truncation of the system is valid: the magnitudes of the expansion coefficients with $n > 11$ are expected to be negligibly small.\footnote{As an additional check, we examined the effect of changing $M$ on our results for the power spectrum.  We found that increasing $M$ from 9 to 11 typically induced a change of $\lesssim 0.01\%$ in $\mathcal{P}_{\mathcal{R}_{c}}$.}

\subsection{Case 3}\label{sec:case 3 numeric}

In this case, we need to numerically solve (\ref{eq:schrodinger 3}), with $\mathcal{A}$ given by (\ref{eq:case 3 A}) and initial data (\ref{eq:case 3 ICs}), in order to obtain the power spectrum.  As in the previous case, it is convenient to introduce a new time coordinate $\tau$, which in this case is defined by
\begin{equation}
	\tau = g + \ln g = -\epsilon k \eta + \ln(-\epsilon k \eta).
\end{equation}
This time parameter allows for good resolution in the superhorizon regime while reducing to $g$ in the early time (subhorizon) limit.  The PDE to solve becomes
\begin{equation}\label{eq:transformed PDE}
	\frac{\di \varphi}{\di \tau} = i\frac{df}{d\tau} \left[  - \frac{1}{2} \frac{\di^{2}}{\di y^{2}} +  V(y,g) \right] \varphi, \quad V(y,g) =  \frac{\tanh^{2}(\sqrt{g} y)}{2g}.
\end{equation}
The analytic boundary conditions are that $\varphi \rightarrow 0$  as $y \rightarrow \pm \infty$.  For our numerical simulations, we impose
\begin{equation}
	\varphi(\tau,\pm\ell) = 0,
\end{equation}
where $\ell$ is a ``sufficiently large'' but finite number.  We introduce a discrete $y$ lattice with $M+2$ nodes as follows:
\begin{equation}
	y_{j} = jh-\ell, \quad h = \frac{2\ell}{M+1}, \quad j = 0 \ldots M+1.
\end{equation}
The values of the wavefunction on this lattice are denoted by
\begin{equation}
	\varphi_{j}(\tau) = \varphi(\tau,y_{j}),  \quad \varphi_{0} = \varphi_{M+1} = 0.
\end{equation}
Employing a centered finite difference stencil $\di_{y}^{2} \varphi \mapsto (\varphi_{j+1} - 2 \varphi_{j} + \varphi_{j-1})/h^{2}$, we can recast the PDE (\ref{eq:transformed PDE}) as a matrix ODE:
\begin{equation}\label{eq:matrix ODE}
	\frac{d \vec{\varphi}}{d\tau} = A \vec{\varphi}, \quad \vec{\varphi} = [\varphi_{1} \cdots \varphi_{M}]^{T}.
\end{equation}
Here, $A=A(\tau)$ is a symmetric, anti-Hermitian and tridiagonal $M \times M$ matrix with
\begin{equation}
	A_{j,j} = i \frac{df}{d\tau} \left[ \frac{1}{h^{2}} + V(y_{j},g) \right], \quad A_{j,j+1} = -i \frac{df}{d\tau} \frac{1}{2h^{2}}
\end{equation}
We solve the matrix ODE (\ref{eq:matrix ODE}) using the same numerical scheme as employed in Appendix \ref{sec:case 2 numeric}:  That is, we introduce a temporal lattice:
\begin{equation}
	\tau_{i} = \tau_{0} + si, \quad \vec{\varphi}_{i} = \vec\varphi(\tau_{i}),
\end{equation}
where $\tau_{0}$ is an initial time and $s$ is the timestep.  Our temporal stencil is
\begin{equation}\label{eq:stencil 3}
	\vec\varphi_{i+1} = U(\tau_{i}) \vec\varphi_{i}, \quad U(\tau_{i}) = \left[I-\tfrac{1}{2} s A(\tau_{i}) \right]^{-1} \left[I+ \tfrac{1}{2} s A(\tau_{i}) \right].
\end{equation}
As above, the evolution matrix is unitary $U^{\dag} U = I$, which means the norm of $\vec\phi$ is conserved: $\vec\varphi_{i}^{\dag} \vec\varphi_{i}= \vec\varphi_{i+1}^{\dag} \vec\varphi_{i+1}$.  Repeated application of the the stencil (\ref{eq:stencil 3}) allows us to evaluate $\varphi$ for $k \ll Ha$ and hence obtain $\langle \chi_{\k}^{2} \rangle$ using (\ref{eq:chi squared}).

\acknowledgments
ACD and SSS are supported by National Sciences and Engineering Research Council of Canada (NSERC).  SSS was also partially supported by the Perimeter Institute for Theoretical Physics' affiliate program.  Research at Perimeter Institute is supported by the Government of Canada through Industry Canada and by the Province of Ontario through the Ministry of Economic Development \& Innovation.  Computational facilities were provided by ACEnet, the regional high performance computing consortium for universities in Atlantic Canada. ACEnet is funded by the Canada Foundation for Innovation (CFI), the Atlantic Canada Opportunities Agency (ACOA), and the provinces of Newfoundland and Labrador, Nova Scotia, and New Brunswick.

\bibliographystyle{apsrev4-1}
\bibliography{DQA_inflation}

\end{document}